%Paper: hep-ph/9312239
%From: peris@surya11.cern.ch (Santiago Peris)
%Date: Tue, 7 Dec 1993 17:24:17 +0100 (MET)

%%%%%%%%%%%%%%%%%%%%%%%%%%%%%%%%%%%%%%%%%%%%%%%%%
% This file is written using phyzzx
%%%%%%%%%%%%%%%%%%%%%%%%%%%%%%%%%%%%%%%%%%%%%%%%%
\input phyzzx

%\physrev

\def\j{\journal}
\def\N{N_c}

\date{CERN-TH.7109/93}
\pubtype{}
\titlepage
\title{HIGHER-ORDER CORRECTIONS TO THE LARGE-$N_c$
BOUND ON $M_{\eta}/M_{\eta '}$}
\author{S. Peris\foot{On leave from Grup de Fisica Teorica and IFAE,
Universitat Autonoma de Barcelona, Barcelona, Spain. e-mail: PERIS @
CERNVM.CERN.CH}}
\address{Theory Division, CERN,\break CH-1211 Geneva 23\break
Switzerland}
\abstract

Next-to-leading $1/\N$ corrections to the upper bound on $M_{\eta}/M_{\eta '}$
recently obtained by Georgi are considered. These corrections are just what
is needed to reconcile
the bound with the observed $\eta$ and $\eta'$ masses.

\vskip 1.5in

CERN-TH.7109

December 1993

\endpage

Recently Georgi\Ref\Georgi{H. Georgi, HUTP-93/A029, hep-ph/9310337}
 has observed an amusing fact concerning the large-$N_c$ approximation to QCD
\Ref\tHooft{G. 't Hooft \j Nucl. Phys. &B72 (74) 461} when applied to the
$\eta$--$\eta '$ system\Ref\Witten{E. Witten\j Nucl. Phys. &B156 (79) 269;
G. Veneziano  \j Nucl. Phys. &B159 (79) 213; P. di Vecchia and G. Veneziano
\j Nucl. Phys. &B171 (80) 253; E. Witten \j Ann. of Phys. &128 (80) 363; C.
Rosenzweig, J. Schechter and C.G. Trahern\j Phys. Rev. &D21 (80) 3388.}.
To lowest nontrivial order in $1/N_c$ and
in the quark masses he
has found that  ${M^2_{\eta}\over M^2_{\eta '}} \leq
 {3-\sqrt{3}\over 3+\sqrt{3}} + {\cal O}(\coeff{m_{u,d}}{m_s})$.
The experimental number is {\it higher}
than this upper bound. Consequently it is mathematically impossible (and
not just inaccurate) to fit the experimentally observed masses within this
approximation.

The purpose of the present short note is to show that higher orders in
$1/N_c$ eliminate this bound. This is of course no surprise since higher
orders means new operators with unknown coefficients, so that the freedom in
parameter space is larger, making it possible to avoid the constraints that
lead to the bound of ref. [\Georgi]. It should be noticed, however, that it
is because the bound in [\Georgi] is very close to the experimental number
that this is possible. Corrections in $1/N_c$, being corrections, should be
``small'', and it is hard to believe that they could fix this problem if
Georgi's bound had turned out to be very different from
the experimental number.

\Ref\Veneziano{P. Di Vecchia, F. Nicodemi, R. Pettorino and G. Veneziano\j
Nucl. Phys. &B181 (81) 318.}

Large-$N_c$ arguments provide a
beautiful explanation of what used to be known as the U(1) puzzle
\refmark{\Witten}; they offer us an interesting way (if not the only one)
to get a handle on the physics of the $\eta '$ from first
principles, i.e. from QCD. It would have been very disturbing if the
bound had remained after higher $1/N_c$ corrections were included, so it was
necessary to check that this indeed does not happen. On
the other hand it was already found in ref. [\Veneziano] that some of the
physics of the $\eta'$, such as the decay $\eta '\to \eta \pi \pi$,
cannot be described, even qualitatively, without
going to higher orders in $1/N_c$.

The limit $\N\to \infty$ offers a consistent way to turn off the anomaly. In
this limit, and in a world of massless quarks, the $\eta '$ truly becomes
the Goldstone boson of the $U(1)_A$ symmetry that is seen at the level of
the QCD Lagrangian. It is then expected that the real world will be
reached from this limit by means
of a combined perturbative expansion in the quark
mass, $m_q$, and $1/\N$.

We shall now show that higher-order corrections in the quark mass
alone cannot reconcile the lowest-order bound obtained by Georgi
with the empirical
masses because this bound persists, in fact, to all orders in $m_q$ in the
limit $\N\to \infty$, and not only to first order as originally derived
in ref. [\Georgi]. Therefore, higher-order $1/\N$ corrections are absolutely
essential for this reconciliation\foot{The following discussion can be
considered as a generalization of the results obtained in ref. [\Georgi ] and
originated from an illuminating comment by G. Veneziano that we gratefully
acknowledge.}.

Let us start with the Lagrangian of QCD for massive quarks
in the limit $\N\to \infty$. Because
there are no OZI violating $q\bar q$ annihilation diagrams in this limit,
there exists a symmetry $U(1)_q\times U(1)_{\bar q}$\ \Ref\Gell{M.
Gell-Mann,
as quoted in G. Veneziano\j Nucl. Phys. &B159 (79) 213.} for each quark
flavor that transforms {\it independently} quarks and antiquarks. Therefore
the mass matrix, ${\cal M}$, for the $\pi^0$
($={u\bar u-d \bar d\over \sqrt 2}$), $\eta_8$
($={u\bar u+d \bar d- 2 s\bar s\over \sqrt 6}$) and $\eta_0$
($={u\bar u+d \bar d+s\bar s\over \sqrt 3}$) mesons must read, in the basis
$(u \bar u, d \bar d, s \bar s)$,

$${\cal M}^2=\pmatrix{A&0&0\cr
                      0&B&0\cr
                      0&0&C\cr}\ \qquad .\eqn\oneone$$
Non-diagonal entries vanish because they originate from
symmetry-violating transitions in which a
quark with a given flavor comes in but does not get out.

If we further take the reasonable limit  $ m_u=m_d=0, m_s\not= 0$, then it
turns out that there exists a further $SU(2)_A\subset SU(2)_L\times SU(2)_R$
symmetry rotating the up and down quarks. Under this symmetry the $\pi^0$
meson gets shifted by a constant amount, proportional to its decay constant,
and
therefore any mass term (i.e. $\pi^0$--$\pi^0$, $\pi^0$--$\eta_8$ and
$\pi^0$--$\eta_0$) must vanish. Hence $A$ and $B$ must be zero and the mass
matrix ${\cal M}$ reads
$${\cal M}^2=\pmatrix{0&0&0\cr
                      0&0&0\cr
                      0&0&C\cr}\ \qquad .\eqn\twotwo$$
If one now wishes
to include the anomaly as the lowest-order correction in $1/\N$, one obtains
$${\cal M}^2=\pmatrix{0&0&0\cr
                      0&0&0\cr
                      0&0&C\cr} +\  {a\over \N}\
             \pmatrix{1&1&1\cr
                      1&1&1\cr
                      1&1&1\cr}\qquad ,\eqn\threethree$$
where $a$ is the parameter that measures the strength of the anomaly. This
mass matrix is exactly of the same form as that of ref.
[\Georgi]\foot{In ref. [\Georgi] $C$ was approximated by its lowest-order
value, i.e.
$C=m_s\times $const.} and leads to the same mass ratio:
$${M^2_{\eta}\over M^2_{\eta'}}={3+R-\sqrt{9-2 R+R^2}\over
3+R+\sqrt{9-2 R+R^2}}\qquad ,\eqn\fourfour$$
with $R\equiv C\N/a$. This mass ratio is maximized for $R=3$ and one
obtains
$${M^2_{\eta}\over M^2_{\eta'}}\leq {3 - \sqrt 3\over 3+\sqrt 3} + {\cal O}
\left( {m_{u,d}\over m_s}\right)\qquad \eqn\fivefive$$
as in ref. [\Georgi]. However, from this derivation we see that
this result is valid to {\it all} orders in
the quark mass in the limit $\N \to \infty$.

The above discussion tells us that consideration of $1/\N$ corrections will
be crucial when discussing modifications to the bound \fivefive . As a matter
of fact what
one has is a combined series expansion in the quark mass and $1/\N$. To
lowest nontrivial order, contributions to the Goldstone boson mass
matrix are due to operators of order $m_q$ and $a/\N$ (i.e. the anomaly). To
next-to-leading order, one must certainly take into account corrections of
order $1/\N$ to the previous operators but also corrections of order
$m_q^2$, at least in principle. Because of the above discussion,
however, contributions that
are quadratic in the quark mass will not affect the bound and may
consequently
be disregarded. This makes the following analysis considerably simpler.

Since we will not deal with any strong CP
violation effect, we shall set $\theta_{QCD}=0$.\Ref\Eduard{For the case
$\theta_{QCD}\not= 0$ one can see, for instance, A. Pich and E. de Rafael\j
Nucl. Phys. &B367 (91) 313 and references therein.} Therefore we
shall next consider the quadratic part of the Lagrangian
describing the $\eta$--$\eta'$ system to next-to-leading order in $1/\N$.
It can be obtained from\refmark{\Veneziano}

$${\cal L}= {\cal L}_0\  +\  \delta{\cal L}$$
$$\eqalign{{\cal L}_0=&{f_{\pi}^2\over 4} \Big[\ \Tr{\partial_{\mu} U
\partial^{\mu} U^{\dag}}\ +\ \Tr{(\chi U^{\dag} + U \chi ^{\dag})}
+\ {a\over 4 \N} \big(\Tr{\log U}-\Tr{\log U^{\dag}}\big)^2 \ \Big]\quad ,\cr
\delta{\cal L}=&{f_{\pi}^2\over 4}\Big[\ {2 \alpha \over 3 \N}
\Tr{U^{\dag}\partial_{\mu}U} \ \Tr{U^{\dag}\partial^{\mu}U}\  + \cr
+ &\ {\epsilon \over 2 \sqrt{2} \N} \big(\Tr{\log U}-\Tr{\log U^{\dag}}\big)\
\Tr{(\chi U^{\dag} - \chi^{\dag} U)}\  \Big] + ...\qquad ,\cr}\ \eqn\one$$
where\foot{We follow the notation of ref. [\Eduard].}
 $U=exp\left({\coeff{-i \sqrt{2} \Phi(x)}{f_{\pi}}}\right)$ with
$\Phi(x)= \coeff{\Phi^0}{\sqrt{3}}+\coeff{\vec \lambda \cdot \vec \Phi}
{\sqrt{2}}$
and $\vec \lambda$ are the eight Gell-Mann matrices, $f_{\pi}\simeq 93$MeV
is the pion decay constant; $\chi= 2 B_0 M$ where
$B_0$ is a parameter related to the quark condensate in QCD and $M$ is the
quark mass matrix. In this Lagrangian $a,\alpha$ and $\epsilon$ are
parameters of ${\cal O} (\N^0)$. Then the first two terms in ${\cal L}_0$ are
of ${\cal O}(p^2 \N^0)$ and the term proportional to $a$ is of
${\cal O}(a\N^{-1})$. The Lagrangian ${\cal L}_0$ is to be considered the
leading-order Lagrangian, and $\delta {\cal L}$ is the $1/\N$
correction to it. Contributions of order
$a/\N^2$ to the mass matrix can be absorbed in a redefinition of $a$.

Using the Lagrangian \one, it is straightforward to compute the mass
matrix in the ($\eta_8$,$\eta_0$) basis. Neglecting terms proportional to
the up and down quark masses but not to the strange quark mass,
one finds
\foot{Notice that the term in eq. \one\ proportional to $\alpha$
affects the mass matrix through the normalization of the kinetic term.}

$$M^2= \ {4\over 3} M_K^2\  \pmatrix{1&-{y\over \sqrt{2}}\cr
                     -{y\over \sqrt{2}}&{y^2\over 2}+ x \cr}\qquad , \eqn\two$$

where

$$y\equiv 1+\delta y={1-{3\epsilon\over \N\sqrt{2}}\over
\sqrt{1-{2\alpha\over \N}}}\approx
1-{1\over \N}\left({3\epsilon\over \sqrt{2}}-\alpha\right)+{\cal O}
\left({1\over \N^2}\right)
\qquad , \eqn\three$$

$$x\equiv {9 a\over 4 M_K^2 \N (1- {2\alpha\over \N})}\approx
{9 a\over 4 M_K^2 \N} \left(1+ {2\alpha\over \N}+{\cal O}
\left({1\over \N^2}\right)\right)
\qquad .\eqn\four$$
Amusingly, although our Lagrangian \one\  has three unknown parameters ($a,
\epsilon$ and $ \alpha$) to start with ($B_0 m$ is fixed through the kaon
mass), the mass matrix \two\ depends only on two combinations of them,
i.e. $x$ and $y$.

In the spirit of the $1/\N$ expansion one should take
$\delta y$ as a small parameter and expand in it. The $\eta$ and $\eta '$
masses are then determined by the conditions
$${3(M^2_{\eta}+M^2_{\eta'})\over 4 M^2_K}={3\over 2} +x+ \delta y\qquad
, \eqn\five$$
$${9 M^2_{\eta} M^2_{\eta '}\over 16 M^4_K}=x\qquad .\eqn\six$$

These equations yield $\delta y \simeq -0.35$
and $x\simeq 2.57$ when the masses
$M_K\simeq 495$MeV, $M_{\eta}\simeq 547$MeV, $M_{\eta'}\simeq 958$MeV
are used. It is clear that the system \five\ + \six\  has always one
solution for $x$ and $\delta y$, once the masses for the pseudoscalars are
given; the mass matrix \two\  {\it can} thus fit the $\eta$ and $\eta '$
masses, and the bound of ref. [\Georgi] is overcome. As a matter of fact,
taking the mass matrix \two\  one easily obtains

$${M^2_{\eta}\over M^2_{\eta'}}=
{1+x+y^2/2-\sqrt{(1+x+y^2/2)^2-4 x}\over
1+x+y^2/2+\sqrt{(1+x+y^2/2)^2-4 x}}\qquad .\eqn\seven $$

This expression has a maximum when varied with respect to $x$ ($\sim$ the
anomaly to quark mass ratio), keeping $y$ fixed. One can understand
this on physical grounds: for $x\to \infty$ the mass ratio \seven\ goes to
zero because the $\eta'$ mass becomes infinite. Furthermore, when $x\to 0$
the mass ratio also goes to zero because the $\eta$ mass vanishes since
the situation of the $U(1)_A$ problem is reproduced. So
there must be an intermediate value of $x$ at which eq. \seven\ has a maximum.
Moreover, in this eq. \seven\  $y$ equals unity only when next-to-leading
terms in $1/\N$ are
neglected. In this case one can
see that eq. \seven\  reaches its maximum at $x_0=3/2$ and one then obtains
the bound\refmark{\Georgi} \fivefive\ . However,
if next-to-leading terms are included  one has instead that
$y\approx 1 + \delta y$ with $\vert \delta y\vert \sim 1/\N <<1$.
Expanding eq. \seven\ in $\delta y$ about $x_0$ one obtains
$${M^2_{\eta}\over M^2_{\eta'}}\leq {3-\sqrt{3}\over 3+\sqrt{3}}
(1- {2\over \sqrt{3}} \delta y) \qquad .\eqn\eight$$
A negative value for $\delta y$ of
order $1/\N\sim 0.3$ is more than enough for eq. \eight\  to be
satisfied experimentally.

\Ref\Narison{G. Veneziano in ref. \Witten; for an update see for instance
S. Narison\j Phys. Lett. &B255 (91) 101.}

\Ref\Donoghue{J.F. Donoghue, B.R. Holstein and Y.-C.R. Lin\j Phys. Rev.
Lett. &55 (85) 2766; J. Gasser and H. Leutwyler \j Nucl. Phys.
&B250 (85) 465; see however G.M. Shore and G. Veneziano\j Nucl. Phys. &B381
(92) 3.}

One must also assess the consistency of the $1/\N$ expansion. First of all,
the size of the $1/\N$ corrections, $\vert \delta y\vert \simeq 0.35$, is
indeed of order $1/\N$. Secondly, within our approximation one finds that
$f_{\eta_0}^2=f^2_{\pi} (1-\coeff{2\alpha}{\N})$ for the decay constants.
The analysis of ref. [\Donoghue] obtains that $f_{\eta_0} \simeq f_{\pi}$,
which would suggest that $\alpha$ is small. Then the result for $x$ leads
essentially to the same value of $a$ and the same
estimate for the topological susceptibility as the original
work of Veneziano \refmark{\Narison}. Notice that this is a consequence of
the fact that eq. \six\  is $y$-independent, which in
turn stems from the particular
$y$-dependence of the mass matrix \two. However, were
we to compute the $\eta$--$\eta'$ mixing angle, we would obtain around
$10^{\circ}$,
i.e. half the experimental number. This is by now a well-known fact that
results from a
fortuitous approximate cancelation in the expression for this angle, which
occurs in lowest non-trivial order
in $1/\N$ and $m_q$. This makes the $\eta$--$\eta'$
mixing angle a very sensitive
parameter whose calculation can be reconciled with experiment
only after corrections of ${\cal O}(m_{q}^2)$ are included, as suggested in
ref. [\Donoghue]. These corrections will also modify the
relation $f_{\pi}=f_{\eta_8}$. However, a full calculation
including next-to-leading
terms in $1/\N$ and terms of ${\cal O}(m_{q}^2)$, although
interesting, is beyond the scope of the present short note.

\ack

I am truly indebted to Toni Pich, Eduard de Rafael and G. Veneziano for very
interesting discussions and for comments on the manuscript. I am also grateful
to S. Narison for conversations.

\refout

\end